\documentclass{llncs}
\usepackage{makeidx}  % allows for indexgeneration
\usepackage[dvips]{epsfig}
\begin{document}
\frontmatter          % for the preliminaries
\pagestyle{headings}  % switches on printing of running heads
\mainmatter              % start of the contributions
\title{Computational Physics on Graphics Processing Units}
\titlerunning{Computational Physics on GPUs}  % abbreviated title (for running head)
\author{Ari Harju\inst{1,2} \and Topi Siro\inst{1,2} \and Filippo Federici~Canova\inst{3} \and Samuli Hakala\inst{1} \and Teemu Rantalaiho\inst{2,4}}
\authorrunning{Ari Harju et al.} % abbreviated author list (for running head)
\institute{COMP Centre of Excellence, Department of Applied Physics, Aalto University School of Science, Helsinki, Finland
%\\
%\email{ari.harju@aalto.fi},\\ WWW home page:
%\texttt{http://physics.aalto.fi}
\and
Helsinki Institute of Physics, Helsinki, Finland\\
\and
Department of Physics, Tampere University of Technology, Tampere, Finland
\and
Department of Physics, University of Helsinki, Helsinki, Finland.
}

\maketitle              % typeset the title of the contribution

\begin{abstract}
The use of graphics processing units for scientific computations is an
emerging strategy that can significantly speed up various 
algorithms. In this review, we discuss advances made in the field of
computational physics, focusing on classical molecular dynamics and
quantum simulations for electronic structure calculations using the
density functional theory, wave function techniques and quantum field theory.
\end{abstract}
\section{Introduction}
The graphics processing unit (GPU) has been an essential part of
personal computers for decades. Their role became much more important
in the 90s when the era of 3D graphics in gaming started. One
of the hallmarks of this is the violent first-person
shooting game DOOM by the id Software company, released in 1993. Wandering around
the halls of slaughter, it was hard to imagine these games leading to
any respectable science. However, twenty years after the
release of DOOM, the gaming industry of today is enormous, and the
continuous need for more realistic visualizations has led to a
situation where modern GPUs have tremendous computational
power. In terms of theoretical peak performance, they have far surpassed the central processing units (CPU).

\begin{figure}[ht]
\begin{center}
\includegraphics[width=.8\columnwidth]{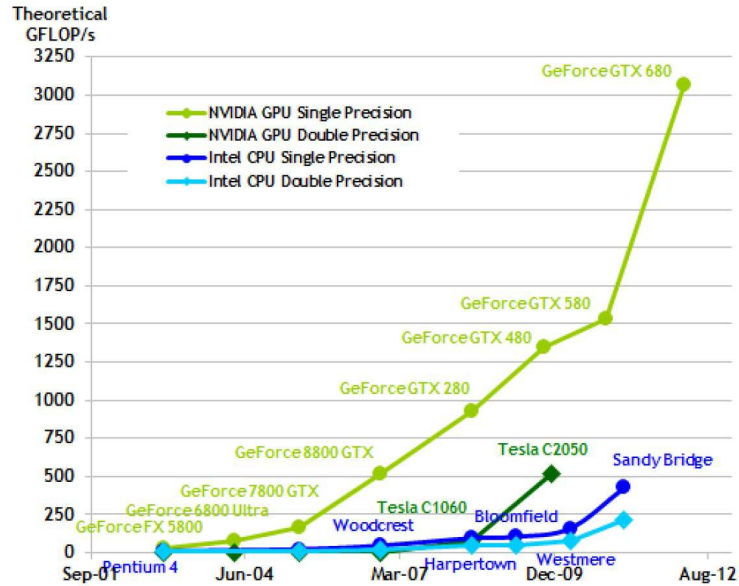}
\end{center}
\caption{Floating point operations (FLOPS) per second for GPUs and
  CPUs from NVIDIA and Intel Corporations, figure taken
  from  \cite{cuda}. The processing power of the currently best GPU
  hardware by the AMD Corporation is comparable to NVIDIA at
  around 2600 GFLOPS/s.}
\label{fig:flops}
\end{figure}

The games started to have real 3D models and hardware acceleration in the mid 90s,
but an important turning point for the scientific use of GPUs for computing was around the 
first years of this
millennium \cite{Macedonia2003106}, when the widespread programmability of GPUs
was introduced. Combined with the continued increase in computational
power as shown in Fig.~\ref{fig:flops}, the GPUs are nowadays a
serious platform for general purpose computing. Also, the memory
bandwidth in GPUs is very impressive. The three main vendors for GPUs,
Intel, NVIDIA, and ATI/AMD, are all actively developing computing on
GPUs. At the moment, none of the technologies listed
above dominate the field, but NVIDIA with its CUDA programming
environment is perhaps the current market leader.

\subsection{The GPU as a computational platform}

At this point, we have hopefully convinced the reader that GPUs feature a
powerful architecture also for general computing, but what makes GPUs
different from the current multi-core CPUs? To understand this, we can
start with traditional graphics processing, where hardware
vendors have tried to maximize the speed at which the pixels on the
screen are calculated. These pixels are independent primitives that
can be processed in parallel, and the number of pixels on computer
displays has increased over the years from the original DOOM
resolution of $320\times 200$, corresponding to 64000 pixels, to
millions. The most efficient way to process these primitives is to
have a very large number of arithmetic logical units (ALUs) that are
able to perform a high number of operations for each video frame. The
processing is very data-parallel, and one can view this as performing the same
arithmetic operation in parallel for each primitive. 
Furthermore, as the operation is the same for each
primitive, there is no need for very sophisticated flow control in the GPU
and more transistors can be used for arithmetics, resulting in an
enormously efficient hardware for performing parallel computing that
can be classified as ``single instruction, multiple data'' (SIMD).

Now, for general computing on the GPU, the primitives are no
longer the pixels on the video stream, but can range from matrix
elements in linear algebra to physics related cases where the
primitives can be particle coordinates in classical molecular dynamics
or quantum field values. Traditional graphics processing teaches us that
the computation would be efficient when we
have a situation where the same calculation needs to be performed for
each member of a large data set. It is clear that not all problems or algorithms have
this structure, but there are luckily many cases where this applies, and the list of successful examples is long.

However, there are also limitations on GPU computing. First of all,
when porting a CPU solution of a given problem to the GPU, one might need
to change the algorithm to suit the SIMD approach. Secondly, the
communication from the host part of the computer to the GPU part is
limited by the speed of the PCIe bus coupling the GPU and the host. In
practice, this means that one needs to perform a serious amount of
computing on the GPU between the data transfers before the GPU can
actually speed up the overall computation. Of course, there are also
cases where the computation as a whole is done on GPU, but these cases
suffer from the somewhat slower serial processing speed of the GPU.

Additional challenges in GPU computing include the often substantial programming effort to get a working and optimized code. While writing
efficient GPU code has become easier due to libraries and programmer friendly hardware features, it still requires some specialized thinking. For example, the
programmer has to be familiar with the different kinds of memory on the GPU to know how and when to use them. Further, things like occupancy of the multiprocessors (essentially, how full the GPU is) and memory access patterns of the threads are something one has to consider to reach optimal performance. Fortunately, each generation
of GPUs has alleviated the trouble of utilizing their full potential. For example, a badly aligned memory access in the first CUDA capable GPUs from NVIDIA could cripple the
performance by drastically reducing the memory bandwidth, while in the Fermi generation GPUs the requirements for memory access coalescing are much more forgiving.

\section{Molecular dynamics}

\begin{figure}
\center
	\includegraphics[width=300 pt]{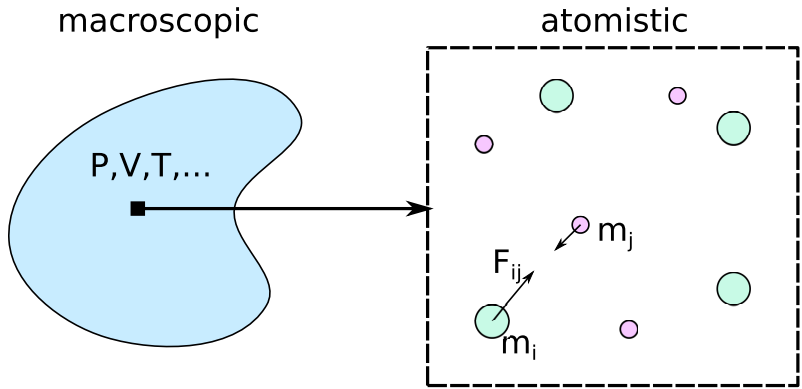}
	\caption{Schematic presentation of the atomistic model of a macroscopic system.}
	\label{fig:md1}
\end{figure}

Particle dynamics simulation, often simply called Molecular dynamics (MD), refers to the type of simulation where the behaviour of a complex system is calculated by integrating the equation of motion of its components within a given model, and its goal is to observe how some ensemble-averaged properties of the system originate from the detailed configuration of its constituent particles (Fig. \ref{fig:md1}). 

In its classical formulation, the dynamics of a system of particles is described by their Newtonian equations:
\begin{equation}
m_i \frac{d^2 \vec{x}_i}{dt^2} = \sum_j \vec{F}_{ij}
\label{eq:newton}
\end{equation}
where $m_i$ is the particle's mass, $\vec{x}_i$ its position, and $\vec{F}_{ij}$ is the interaction between the $i$-th and $j$-th particles as provided by the model chosen for the system under study. These second order differential equations are then discretised in the time domain, and integrated step by step until a convergence criterion is satisfied.

The principles behind MD are so simple and general that since its first appearance in the 70s, it has been applied to a wide range of systems, at very different scales. For example, MD is the dominant theoretical tool of investigation in the field of biophysics, where structural changes in proteins \cite{McC77,Tem84,Gao89,Sam11} and lipid bilayers 
 \cite{Ber10,Lyu11} interacting with drugs can be studied, ultimately providing a better understanding of drug delivery mechanisms. 

At larger scales, one of the most famous examples is known as the \emph{Millenium Simulation}, where the dynamics of the mass distribution of the universe at the age of 380000 years was simulated up to the present day \cite{Spr05nature}, giving an estimate of the age of cosmic objects such as galaxies, black holes and quasars, greatly improving our understanding of cosmological models and providing a theoretical comparison to satellite measurements.

Despite the simplicity and elegance of its formulation, MD is not a computationally easy task and often requires special infrastructure. The main issue is usually the evaluation of all the interactions $\vec{F}_{ij}$, which is the most time consuming procedure of any MD calculation for large systems. Moreover, the processes under study might have long characteristic time scales, requiring longer simulation time and larger data storage; classical dynamics is chaotic, i.e. the outcome is affected by the initial conditions, and since these are in principle unknown and chosen at random, some particular processes of interest might not occur just because of the specific choice, and the simulation should be repeated several times. For these reasons, it is important to optimise the evaluation of the forces as much as possible.

%Fortunately, MD is a very suitable problem for parallel computation, as all the interactions $\vec{F}_{ij}$ are independent of each other and their evaluation can be distributed over different computing units; custom hardware architectures dedicated to a specific MD simulation were also developed \cite{Bak90,Sus03,Nis09}, and despite the good performance achieved, their excessive cost and lack of general applicability hampered their spread through scientific communities, or even beyond the particular scope they were initially intended for.

%Recently, GPUs have attracted a lot of interest from the scientific community because they feature a massively parallel architecture with performance on the level of small computer clusters at the cost and power consumption of conventional commodity hardware. Even though they were originally designed for real-time 3D rendering, the computer games market drove GPU manufacturers to fit increasingly more processing power on their devices, to the point where it became advantageous to calculate the game physics (dynamics and collisions) on the device itself \cite{gpu07rigidb,gpu07coll}. 

An early attempt to implement MD on the GPU was proposed in 2004 \cite{Chi04} and showed promising performance; at that time, general purpose GPU computing was not yet a well established framework and the N-body problem had to be formulated as a rendering task: a \emph{shader} program computed each pair interaction $\vec{F}_{ij}$ and stored them as the pixel color values (RBG) in an $N\times N$ texture. Then, another shader would simply sum these values row-wise to obtain the total force on each particle and finally integrate their velocities and positions. The method is called all-pairs calculation, and as the name might suggest, it is quite expensive as it requires $\mathcal{O}(N^2)$ force evaluations. The proposed implementation was in no way optimal since the measured performance was about a tenth of the nominal value of the device, and it immediately revealed one of the main issues of the architecture that still persists nowadays: GPUs can have a processing power exceeding the teraflop, but, at the same time, they are extremely slow at handling the data to process since a memory read can require hundreds of clock cycles. The reason for the bad performance was in fact the large amount of memory read instructions compared to the amount of computation effectively performed on the fetched data, but despite this limitation, the code still outperformed a CPU by a factor of 8 because every interaction was computed concurrently. A wide overview of optimisation strategies to get around the memory latency issues can be found in Ref. \cite{Bro12}, while, for the less eager to get their hands dirty, a review of available MD software packages is included in Ref. \cite{Sto10}. 

In the current GPU programming model, the computation is distributed in different threads, grouped together as blocks in a grid fashion, and they are allowed to share data and synchronise throughout the same block; the hardware also offers one or two levels of cache to enhance data reuse, thus reducing the amount of memory accesses, without harassing the programmer with manual pre-fetching. A more recent implementation of the all-pair calculation \cite{gpu07nbody} exploiting the full power of the GPU can achieve a performance close to the nominal values, comparable to several CPU nodes. 

The present and more mature GPGPU framework allows for more elaborate kernels to fit in the device, enabling the implementation of computational tricks developed during the early days of MD \cite{allen2002_mdbook} that make it possible to integrate N-body dynamics accurately with much better scaling than $\mathcal{O}(N^2)$.
\begin{figure}
\center
	\includegraphics[width=300 pt]{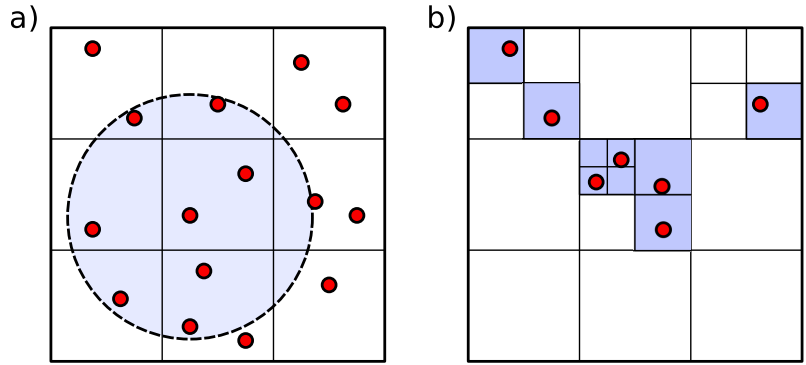}
	\caption{Illustration of different space partition methods. In dense systems (a) a regular grid is preferred, and neighbouring particles can be searched in only a few adjacent voxels. Sparse systems (b) are better described by hierarchical trees, excluding empty regions from the computation.}
	\label{fig:partition}
\end{figure}
For example, in many cases the inter-particle forces are short range, and it would be unnecessary to evaluate every single interaction $\vec{F}_{ij}$ since quite many of them would be close to zero and just be neglected. It is good practice to build lists of neighbours for each particle in order to speed up the calculation of forces: this also takes an $\mathcal{O}(N^2)$ operation, although the list is usually only recalculated every 100-1000 timesteps, depending on the average mobility of the particles. The optimal way to build neighbour lists is to divide the simulation box in voxels and search for a partcle's neighbours only within the adjacent voxels (Fig. \ref{fig:partition}a), as this procedure requires only $\mathcal{O}(N)$ instructions. Performance can be further improved by sorting particles depending on the index of the voxel they belong, making neighbouring particles in space, to a degree, close in memory, thus increasing coalescence and cache hit rate on GPU systems;
  such a task can be done with radix count sort \cite{Kip04,Sat08radix,Ha09} in $\mathcal{O}(N)$ with excellent performance, and it was shown to be the winning strategy \cite{And08}. %Decomposing the simulation space in regular voxels is mostly useful when the system is homogeneous so that the workload is evenly distributed among the thread blocks, but may not be so advantageous for sparse systems, where hierarchical methods are preferred.

Unfortunately, most often the inter-particle interactions are not exclusively short range and can be significant even at larger distances (electrostatic and gravitational forces). Therefore, introducing an interaction cut-off leads to the wrong dynamics.
For dense systems, such as bulk crystals or liquids, the electrostatic interaction length largely exceeds the size of the simulation space, and in principle one would have to include the contributions from several periodic images of the system, although their sum is not always convergent. The preferred approach consists of calculating the electrostatic potential $V(\vec{r})$ generated by the distribution of point charges $\rho(\vec{r})$ from Poisson's equation:
\begin{equation}
\nabla^2 V(\vec{r}) = \rho (\vec{r})
\label{eq:poisson}
\end{equation}
The electrostatic potential can be calculated by discretising the charge distribution on a grid, and solving Eq. \ref{eq:poisson} with a fast Fourier transform (FFT), which has $\mathcal{O}(M log M)$ complexity (where $M$ is the amount of grid points): this approach is called particle-mesh Ewald (PME).
Despite being heavily non-local, much work has been done to improve the FFT algorithm and make it cache efficient \cite{Mor03fft,Gov07,Gu10,Che10,Ahm11}, so it is possible to achieve a 20-fold speed up over the standard CPU FFTW or a 5-fold speedup when compared to a highly optimised MKL implementation.
The more recent multilevel summation method (MSM) \cite{Ske02} uses nested interpolations of progressive smoothing of the electrostatic potential on lattices with different resolutions, offering a good approximation of the electrostatic $\mathcal{O}(N^2)$ problem in just $\mathcal{O}(N)$ operations. The advantage of this approach is the simplicity of its parallel implementation, since it requires less memory communication among the nodes, which leads to a better scaling than the FFT calculation in PME. The GPU implementation of this method gave a 25-fold speedup over the single CPU \cite{Har09}.
Equation \ref{eq:poisson} can also be translated into a linear algebra problem using finite differences, and solved iteratively on multi-grids \cite{Goo05,McA10} in theoretically $\mathcal{O}(M)$ operations. Even though the method initially requires several iterations to converge, the solution does not change much in one MD step and can be used as a starting point in the following step, which in turn will take much fewer iterations. 

On the other hand, for sparse systems such as stars in cosmological simulations, decomposing the computational domain in regular boxes can be quite harmful because most of the voxels will be empty and some computing power and memory is wasted there. The optimal way to deal with such a situation is to subdivide the space hierarchically with an octree \cite{Ren80} (Fig. \ref{fig:partition}b), where only the subregions containing particles are further divided and stored. Finding neighbouring particles can be done via a traversal of the tree in $\mathcal{O}(N log N)$ operations.
Octrees are conventionally implemented on the CPU as dynamical data structures where every node contains reference pointers to its parent and children, and possibly information regarding its position and content. This method is not particularly GPU friendly since the data is scattered in memory as well as in the simulation space. In GPU implementations, the non-empty nodes are stored as consecutive elements in an array or texture, and they include the indices of the children nodes \cite{gpu05octree}. They were proved to give a good acceleration in solving the N-body 
problem \cite{gpu07nbody,Bel08,Ham09}. 
Long range interactions are then calculated explicitly for the near neighbours, while the fast multipole method (FMM) \cite{Rok85,Gre87} can be used to evaluate contributions from distant particles. The advantage of representing the system with an octree becomes now more evident: there exists a tree node containing a collection of distant particles, which can be treated as a single multipole leading to an overall complexity $\mathcal{O}(N)$. Although the mathematics required by FMM is quite intensive to evaluate, the algorithms involved have been developed and extensively optimised for the GPU architecture \cite{Gum08,Dar11,Tak12}, achieving excellent parallel performance even on large clusters \cite{Yok12}.

In all the examples shown here, the GPU implementation of the method outperformed its CPU counterpart: in many cases the speedup is only 4-5 fold when compared to a highly optimised CPU code, which seems, in a way, a discouraging result, because implementing an efficient GPU algorithm is quite a difficult task, requiring knowledge of the target hardware, and the programming model is not as intuitive as for a regular CPU. To a degree, the very same is true for CPU programming, where taking into account cache size, network layout, and details of shared/distributed memory of the target machine when designing a code leads to higher performance. These implementation difficulties could be eased by developing better compilers, that check how memory is effectively accessed and provide higher levels of GPU optimisation on older CPU codes automatically, hiding the complexity of the hardware specification from the programmer.
In some cases, up to 100 fold speedups were measured, suggesting that the GPU is far superior. These cases might be unrealistic since the nominal peak performance of a GPU is around 5 times bigger than that of a CPU. Therefore, it is possible that the benchmark is done against a poorly optimised CPU code, and the speedup is exaggerated. On the other hand, GPUs were also proven to give good scaling in MPI parallel calculations, as shown in Refs. \cite{Ham09} and \cite{Yok12}. In particular, the AMBER code was extensively benchmarked in Ref. \cite{Got12}, and it was shown how just a few GPUs (and even just one) can outperform the same code running on 1024 CPU cores: the weight of the communication between nodes exceeds the benefit of having additional CPU cores, while the few GPUs do not suffer from this latency and can deliver better performance, although the size of the computable system becomes limited by the available GPU memory. It has to be noted how GPU solutions, even offering a modest 4-5 fold speedup, do so at a lower hardware and running cost than the equivalent in CPUs, and this will surely make them more appealing in the future.
From the wide range of examples in computational physics, it is clear that the GPU architecture is well suited for a defined group of problems, such as certain procedures required in MD, while it fails for others. This point is quite similar to the everlasting dispute between raytracing and raster graphics: the former can explicitly calculate photorealistic images in complex scenes, taking its time (CPU), while the latter resorts to every trick in the book to get a visually "alright" result as fast as possible (GPU). It would be best to use both methods to calculate what they are good for, and this sets a clear view of the future hardware required for scientific computing, where both simple vector-like processors and larger CPU cores could access the same memory resources, avoiding data transfer.

\section{Density-functional theory}

Density functional theory (DFT) is a popular method for {\it ab-initio}
electronic structure calculations in material physics and quantum
chemistry. 
In the most commonly used DFT formulation by Kohn and Sham \cite{PhysRev.140.A1133},
the problem of $N$ interacting electrons is mapped to one with $N$ non-interacting electrons
moving in an effective potential so that the total electron density is the same as in the original many-body case \cite{parr1994density}.
To be more specific, the single-particle Kohn-Sham orbitals
$\psi_n(\mathbf{r})$ are solutions to the equation
\begin{equation}
  H\psi_n(\mathbf{r})=\epsilon_n\psi_n(\mathbf{r}),
\end{equation}
where the effective Hamiltonian in atomic units is 
$H=-\frac{1}{2}\nabla^2+v_H(\mathbf{r})+v_{ext}(\mathbf{r})+v_{xc}(\mathbf{r})$.
The three last terms in the Hamiltonian define the effective potential, consisting of
the Hartree potential $v_H$ defined by the Poisson equation
$\nabla^2v_H(\mathbf{r})=-4\pi \rho(\mathbf{r})$, 
the external ionic potential $v_{ext}$, and the exchange-correlation
potential $v_{xc}$ that contains all the complicated many-body physics
the Kohn-Sham formulation partially hides. In practice, the $v_{xc}$ part needs to be
approximated.
The electronic charge density $\rho(\mathbf{r})$ is
determined by the Kohn-Sham orbitals as
$\rho(\mathbf{r})=\sum_if_i|\psi_i(\mathbf{r})|^2$, where the $f_i$:s are
the orbital occupation numbers.

There are several numerical approaches and approximations for solving
the Kohn-Sham equations.  They relate usually to the discretization of the
equations and the treatment of the core electrons (pseudo-potential
and all electron methods). The most common discretization methods in
solid state physics are plane waves, localized orbitals, real space
grids and finite elements. Normally, an iterative procedure called
self-consistent field (SCF) calculation is used to find the solution
to the eigenproblem starting from an initial guess for the charge
density \cite{RevModPhys.64.1045}.

Porting an existing DFT code to GPUs generally includes profiling or
discovering with some other method the computationally most expensive
parts of the SCF loop and reimplementing them with GPUs. Depending on the
discretization methods, the known numerical bottlenecks are vector
operations, matrix products, Fast Fourier Transforms (FFTs) and
stencil operations. There are GPU versions of many of the standard
computational libraries (like CUBLAS for BLAS and CUFFT for FFTW).
However, porting a DFT application is not as simple as replacing the
calls to standard libraries with GPU equivalents since the resulting
intermediate data usually gets reused by non standard and less
computationally intensive routines. Attaining high performance on a
GPU and minimizing the slow transfers between the host and the device
requires writing custom kernels and also porting a lot of the
non-intensive routines to the GPU.

Gaussian basis functions are a popular choice in quantum chemistry to
investigate electronic structures and their properties. They are used
in both DFT and Hartree-Fock calculations. The known computational
bottlenecks are the evaluation of the two-electron repulsion integrals (ERIs)
and the calculation of the exchange-correlation potential. Yasuda was
the first to use GPUs in the calculation of the exchange-correlation
term \cite{doi:10.1021/ct8001046} and in the evaluation of the Coulomb
potential \cite{JCC:JCC20779}. The most complete work in this area was
done by Ufimtsev {\it et al.}. They have used GPUs in
ERIs \cite{4653202,doi:10.1021/ct700268q,doi:10.1021/ct100701w}, in
complete SCF calculations \cite{doi:10.1021/ct800526s} and in energy
gradients \cite{doi:10.1021/ct9003004}. Compared to the mature GAMESS
quantum chemistry package running on CPUs, they were able to achieve
speedups of more than 100 using mixed precision arithmetic in
HF SCF calculations. Asadchev {\it et al.}. have also done an ERI implementation
on GPUs using the uncontracted Rys quadrature
algorithm \cite{doi:10.1021/ct9005079}.

The first complete DFT code on GPUs for solid state physics
was presented by Genovese {\it et al.}. \cite{genovese:034103}. They used
double precision arithmetic and a Daubechies wavelet based code called
BIGDFT \cite{genovese:014109}. The basic 3D operations for a wavelet
based code are based on convolutions. They achieved speedups of
factor 20 for some of these operations on a GPU, and a factor of 6 for
the whole hybrid code using NVIDIA Tesla S1070 cards.  These results
were obtained on a 12-node hybrid machine.

For solid state physics, plane wave basis sets are the most common
choice. The computational schemes rely heavily on linear algebra
operations and fast Fourier transforms. The Vienna ab initio Simulation
Package (VASP) \cite{PhysRevB.54.11169} is a popular code combining
plane waves with the projector augmented wave method. The most time
consuming part of optimizing the wave functions given the trial
wave functions and related routines have been ported to GPUs.
Speedups of a factor between 3 and 8 for the blocked Davinson
scheme \cite{Maintz20111421} and for the RMM-DIIS
algorithm \cite{JCC:JCC23096} were achieved in real-world examples with
Fermi C2070 cards.  Parallel scalability with 16 GPUs was similar to
16 CPUs.  Additionally, Hutchinson {\it et al.}  have done an implementation
of exact-exchange calculations on GPUs for
VASP \cite{Hutchinson20121422}.

Quantum ESPRESSO \cite{0953-8984-21-39-395502} is a electronic
structure code based on plane wave basis sets and pseudo-potentials (PP).
For the GPU version \cite{girotto12:_enabl_quant_espres}, the most
computationally expensive parts of the SCF cycle were gradually
transferred to run on GPUs. FFTs were accelerated by CUFFT, LAPACK by
MAGMA and other routines were replaced by CUDA kernels. GEMM
operations were replaced by the parallel hybrid phiGEMM \cite{6169574}
library. For single node test systems, running with NVIDIA Tesla C2050,
speedups between 5.5 and 7.8 were achieved and for a 32 node parallel
system speedups between 2.5 and 3.5 were observed. Wand et
al. \cite{Wang:2011:LSP:2063384.2063479} and Jia {\it et al.}. \cite{Jia20139}
have done an implementation for GPU clusters of a plane wave
pseudo-potential code called PEtot. They were able to achieve
speedups of 13 to 22 and parallel
scalability up to 256 CPU-GPU computing units.

GPAW \cite{0953-8984-22-25-253202} is a density-functional theory (DFT)
electronic structure program package based on the real space grid
based projector augmented wave method.  We have used GPUs to
speed up most of the computationally intensive parts of the code:
solving the Poisson equation, iterative refinement of the eigenvectors,
subspace diagonalization and orthonormalization of the wave functions.
Overall, we have achieved speedups of up to 15 on large systems and
a good parallel scalability with up to 200 GPUs using NVIDIA Tesla
M2070 cards \cite{hakala12}.

Octopus \cite{PSSB:PSSB200642067,0953-8984-24-23-233202} is a DFT code
with an emphasis on the time-dependent density-functional theory
(TDDFT) using real space grids and pseudo-potentials. Their GPU
version uses blocks of Kohn-Sham orbitals as basic data units. Octopus
uses GPUs to accelerate both time-propagation and ground state
calculations. Finally, we would like to mention the linear
response Tamm-Dancoff TDDFT implementation \cite{doi:10.1021/ct200030k}
done for the GPU-based TeraChem code.

\section{Quantum field theory}

Quantum field theories are currently our best models for fundamental interactions of the natural world (for a brief introduction to quantum field theories -- or QFTs -- see for example \cite{peskinschroeder} or 
 \cite{Crewther:1995wq} and references therein).
Common computational techniques include perturbation theory, 
which works well in quantum field theories as long as the couplings are small enough to be considered as perturbations 
to the free theory. Therefore, perturbation theory is the primary tool used in pure QED, weak nuclear force and high momentum-transfer QCD
phenomena, but it breaks up when the coupling constant of the theory (the measure of the interaction strength) becomes large, such 
as in low-energy QCD. 

Formulating the quantum field theory on a space-time lattice provides an opportunity to study the model non-perturbatively 
and use computer simulations to get results for a wide range of phenomena -- it enables, for example, one to compute the hadronic spectrum
of QCD (see \cite{Fodor:2012gf} and references therein) from first principles and provides solutions for many vital gaps left 
by the perturbation theory, such as structure functions of
composite particles \cite{Gockeler:2004vx}, form-factors \cite{Alexandrou:2012mz} and decay-constants \cite{McNeile:2011ng}. 
It also enables one to study and test models for new physics, such as technicolor theories \cite{Rummukainen:2011xv} and 
quantum field theories at finite temperature \cite{0954-3899-35-4-044033}, \cite{Petreczky:2009ip} or \cite{2009arXiv0908.3341F}. 
For an introduction to Lattice QFT, see for example \cite{Montvay:QuantumFieldsOnALattice},
 \cite{Rothe:IntroGaugeLFT} or \cite{Gupta:1997nd}.

Simulating quantum field theories using GPUs is not a completely new idea and early adopters even used OpenGL (graphics processing library)
to program the GPUs to solve lattice QCD \cite{Egri:2006zm}. The early GPGPU programmers needed to set up a program that draws two triangles that
fill the output texture of desired size by running a ``shader program'' that does the actual computation
for each output pixel. In this program, the input data could
be then accessed by fetching pixels' input texture(s) using the texture units of the GPU. In lattice QFT, where one typically needs to fetch the
nearest neighbor lattice site values, this actually results in good performance as the texture caches and layouts of the GPUs have been optimized
for local access patterns for filtering purposes.

\subsection{Solving QFTs numerically}

\begin{figure}[ht]
\begin{center}
\includegraphics[width=.8\columnwidth]{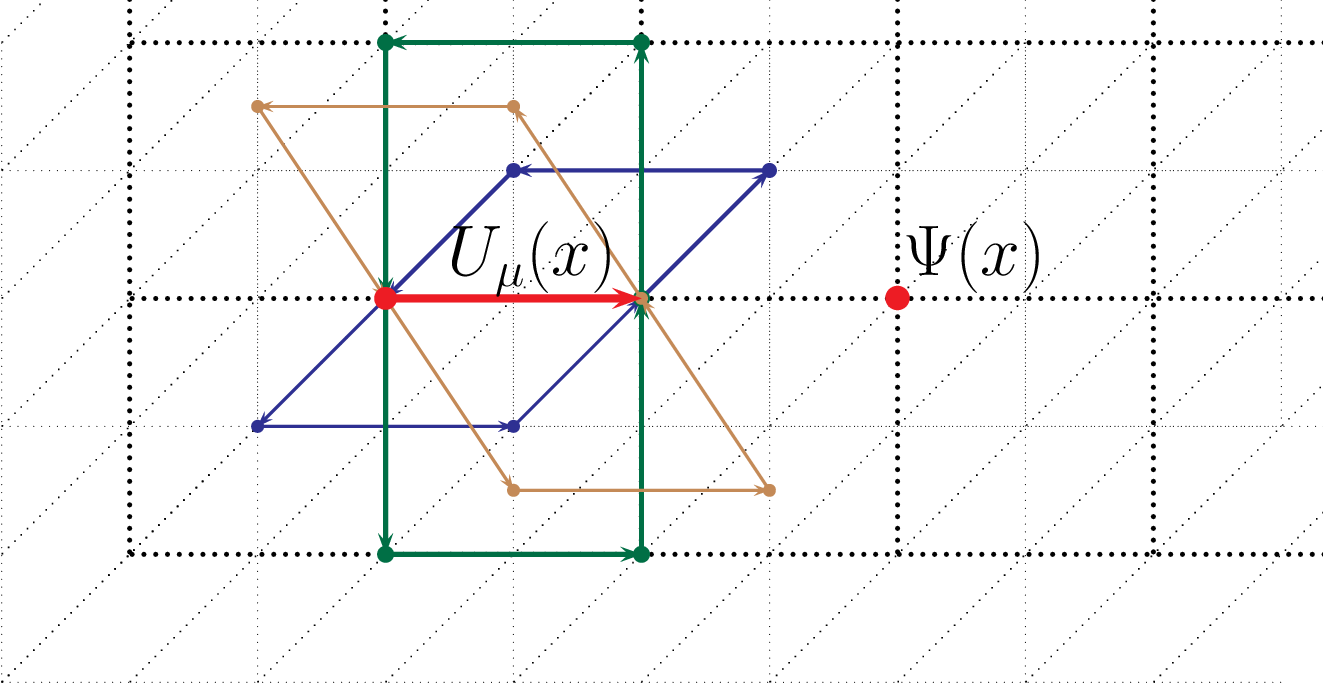}
\end{center}
\caption{The matter fields $\Psi(x)$ live on lattice sites, whereas the gauge fields $U_\mu(x)$ live on the links connecting the sites. Also
depicted are the staples connecting to a single link variable that are needed in the computation of the gauge field forces.}
\label{fig:lattice}
\end{figure}

The idea behind lattice QFT is based on the discretization of the path integral solution to expectation values
of time-ordered operators in quantum field theories. 
First, one divides spacetime into discrete boxes, called the lattice, and 
places the fields onto the lattice sites and onto the links between the sites, as shown in Fig.~\ref{fig:lattice}. Then, one can simulate
nature by creating a set of multiple field configurations, called an \emph{ensemble}, and calculate the values of physical
observables by computing ensemble averages over these states. 

The set of states is normally produced with the help of a Markov chain and
in the most widely studied QFT, the lattice QCD, the chain is produced by combining a \emph{molecular dynamics} algorithm
together with a \emph{Metropolis} acceptance test. Therefore, the typical computational tasks in lattice QFTs are:
 \begin{enumerate}
  \item Refresh generalized momentum variables from a heat bath (Gaussian distribution) 
	  once per \emph{trajectory}.
  \item Compute generalized forces for fields for each step
  \item Integrate classical equations of motion for the fields at each step \footnote{The integration is \emph{not} done
 	with respect to normal time variable, but through the Markov chain index-``time''.}
  \item Perform a Metropolis acceptance test at the end of the trajectory in order to achieve the correct limiting distribution.
 \end{enumerate}
In order to reach satisfying statistics, normally thousands of these trajectories need to be generated and
each trajectory is typically composed of 10 to 100 steps.
The force calculation normally involves a matrix inversion, where the matrix indices run over the entire
lattice and it is therefore the heaviest part of the computation. The matrix arises in simulations with dynamical fermions
(normal propagating matter particles) and the simplest form for the fermion matrix is\footnote{There are multiple different
algorithms for simulating fermions, here we present the simplest one for illustrative purposes.}
\begin{displaymath}
A_{x,y} = [{Q}^{\dagger}Q]_{x,y} \qquad \textrm{where} \quad
\end{displaymath}
\begin{equation} \label{eq:fm}
 Q_{x,y} = \delta_{x,y} - \kappa \sum_{\mu = \pm 1}^{\pm 4} \delta_{y + \hat{\mu}, x}  (1 + \gamma_\mu) U_\mu(x).
\end{equation}
Here, $\kappa$ is a constant related to the mass(es) of the quark(s), $\delta_{x,y}$ is the \emph{Kronecker delta function} (unit matrix elements),
the sum goes over the spacetime dimensions $\mu$, $\gamma_\mu$ are 4-by-4 constant matrices and $U_\mu(x)$ are the link variable matrices that 
carry the force (gluons for example) from one lattice site to the neighbouring one. In normal QCD they are 3-by-3 complex matrices.

The matrix $A$ in the equation $Ar=z$, where one
solves for the vector $r$ with a given $z$, is an almost diagonal sparse matrix with a \emph{predefined sparsity pattern}. 
This fact makes
lattice QCD ideal for parallelization, as the amount work done by each site
is constant.
The actual algorithm used in the matrix inversion is normally some variant of the conjugate gradient algorithm, 
and therefore one needs fast code
to handle the multiplication of a fermion vector by the fermion matrix.

This procedure is the generation of the lattice configurations which form the ensemble. Once the set of configurations $\{U_i\}, i \in [1,N]$ has 
been generated with the statistical weight $e^{-S[U_i]}$, where $S[U_i]$ is the \emph{Euclidean action} (action in imaginary time formulation), the 
expectation value of an operator $F[U]$ can be computed simply as 
\begin{equation}
  \langle F\big[U \big]\rangle \approx 
	  \frac{1}{N}\sum_{i=1}^N F[U_i],
\end{equation}

%\subsection{Existing GPU Solutions to Lattice QFTs} \label{sec:qcd:existing}

As lattice QFTs are normally easily parallelizable, they fit well into the GPU programming paradigm, which can be characterized
as parallel throughput computation. The conjugate gradient methods perform many fermion matrix vector multiplications whose arithmetic
intensity (ratio of floating point operations done per byte of memory fetched) is quite low, making memory bandwidth the normal bottleneck
within a single processor. Parallelization between processors is done by standard MPI domain decomposition techniques. 
The conventional wisdom that this helps due to higher local volume to communication surface area ratio is actually flawed, as typically the GPU
can handle a larger volume in the same amount of time, hence requiring the MPI-implementation to also
take care of a larger surface area in the same time as with a CPU. 
In our experience, GPU adoption is still in some sense in its infancy, as the network implementation 
seems to quickly become the bottleneck in the computation and the MPI implementations of 
running systems seem to have been tailored to meet the needs of the CPUs of the system. Another aspect of this
is that normally the GPUs are coupled with highly powered CPUs in order to cater for the situation where the users
use the GPUs in just a small part of the program and need a lot of sequential performance in order to try to keep the serial
part of the program up with the parallel part.
The GPU also needs a lot of concurrent threads 
(in the order of thousands) to be filled completely with work and therefore good performance is only achievable with relatively large local
lattice sizes. 

Typical implementations assign one GPU thread per site, which makes parallelization easy and gives the compiler quite a lot of room to 
find instruction level parallelism, but in our experience this
can result in a relatively high register pressure: the quantum fields living on the sites have many indices (normally color and Dirac indices)
and are therefore vectors or matrices with up to 12 complex numbers per field per site in the case of quark fields in normal QCD.
Higher parallelization can be achieved by taking advantage of the vector-like parallelism inside a single lattice site, but
this may be challenging to implement in those loops where the threads within a site have to collaborate to produce a result, especially
because GPUs impose restrictions on the memory layout of the fields (consecutive threads have to read consecutive memory locations
in order to reach optimal performance \cite{cuda}). 
In a recent paper \cite{Schrock:2012rm}, the authors solve the \emph{gauge fixing} problem by using overrelaxation techniques and they 
report an increase in performance by using multiple threads per site, although in this case the register pressure problem is even more pronounced
and the effects of register spilling to the L1 cache were not studied. 

The lattice QCD community has a history of taking advantage of computing solutions outside the mainstream: the QCDSP \cite{QCDSP} 
computer was a custom machine that used digital signal processors to solve QCD with an order of one teraflop of performance. 
QCDOC \cite{QCDOC} used a custom IMB powerPC-based ASIC and a multidimensional torus network, which later on evolved into the first version of 
the Blue Gene supercomputers \cite{BLUEGENEL}. The APE collaboration has a long history of custom solutions for lattice QCD and 
is building custom network solutions for lattice QCD \cite{APENETPLUS}. For example, QCDPAX \cite{QCDPAX} was a very early parallel architecture used to study 
Lattice QCD without dynamical fermions.

Currently, there are various groups using GPUs to do lattice QFT simulations. 
The first results using GPUs were produced as early as 2006 in a study that determined the
transition temperature of QCD \cite{2006PhLB..643...46A}.  
Standardization efforts for high precision Lattice QCD libraries are underway and the QUDA library \cite{Babich:2011np} scales to hundreds of GPUs
by using a local Schwarz preconditioning technique, effectively eliminating all the GPU-based MPI communications for a significant portion
of the calculation. They employ various optimization techniques, such as \emph{mixed-precision} solvers, where parts of the inversion process of
the fermion matrix is done at lower precision of floating point arithmetic and using reduced representations of the SU3 matrices. 
Scaling to multiple GPUs can also
be improved algorithmically: already a simple (almost standard) \emph{clover improvement} \cite{Hasenbusch:2002wi} 
term in the fermion action leads to better locality and
of course improves the action of the model as well, taking the lattice formulation closer to the continuum limit. 
Domain decomposition and taking advantage of restricted additive Schwarz (RAS) preconditioning using GPUs was already studied in 2010 in 
 \cite{Osaki:2010v},
where the authors get the best performance on a $32^4$ lattice with vanishing overlap between the preconditioning 
domains and three complete RAS iterations each containing
just five iterations to solve the local system of $4 \times 32^3$ sites.
It should be noted though that the hardware they used is already old, so optimal parameters
with up-to-date components could slightly differ.

Very soon after starting to work with GPUs on lattice QFTs, one notices the effects of Amdahl's law which just points out the fact that 
there is an upper bound for the whole program performance improvement related to optimizing just a portion of the program. It is quite possible
that the fermion matrix inversion takes up 90\% of the total computing time, but making this portion of the code run 10 times faster reveals something odd:
now we are spending half of our time computing forces and doing auxiliary computations and if we optimize this portion of the code as well,
we improve our performance by a factor of almost two again -- therefore optimizing only the matrix inversion gives us a mere fivefold performance
improvement instead of the promised order of magnitude improvement.
Authors of \cite{Bonati:2011dv} implemented practically the entire HMC trajectory on the GPU to fight Amdahl's law and 
recent work \cite{Winter:2011an} on the QDP++ library implements \emph{Just-in-Time} compilation to create GPU kernels on the fly
to accommodate any non-performance critical operation over the entire lattice.

Work outside of standard Lattice QCD using GPUs includes the implementation of the Neuberger-Dirac overlap operator \cite{Walk:2010ut}, which
provides \emph{chiral symmetry} at the expense of a non-local action. Another group uses the Arnoldi algorithm on a multi-GPU cluster
to solve the overlap operator \cite{Alexandru:2011sc} and shows scaling up to 32 GPUs.
Quenched SU2 \cite{Cardoso:2010di} and later quenched SU2, SU3 and generic SU($N_c$) simulations using GPUs 
are described in \cite{Cardoso:2011xu} and 
even compact U(1) Polyakov loops using GPUs are studied in \cite{Amado:2012wt}. 
Scalar field theory -- the so-called $\lambda \phi^4$ model --
using AMD GPUs is studied in \cite{Bordag:2012nh}. 
The TWQCD collaboration has also implemented almost the entire HMC trajectory computation with dynamical Optimal Domain Wall Fermions, which 
improve the chiral symmetry of the action \cite{Chiu:2011bm}.

While most of the groups use exclusively NVIDIA's CUDA-implementation \cite{cuda}, which offers good reliability, flexibility and stability,
there are also some groups using the OpenCL standard \cite{opencl}. A recent study \cite{Bach:2012iw} showed better performance on AMD GPUs
than on NVIDIA ones using OpenCL, although it should be noted that the NVIDIA GPUs were consumer variants with reduced double precision 
throughput and that optimization was done for AMD GPUs. The authors of \cite{Bonati:2011dv} have implemented both CUDA and OpenCL versions of their
staggered fermions code and they report a slightly higher performance for CUDA and for NVIDIA cards.

\subsection{QFT Summary}

All in all, lattice QFT using GPUs is turning from being a promising technology to a very viable alternative to traditional CPU-based computing.
When reaching for the very best strong scaling 
performance -- meaning best performance for small lattices -- single threaded performance does matter if
we assume that the rest of the system scales to remove other bottlenecks (communication, memory bandwith.) 
In these cases, it seems that currently
the best performance is achievable through high-end supercomputers, such as 
the IBM Blue Gene/Q \cite{bluegeneq}, where the microprocessor architecture
is actually starting to resemble more a GPU than a traditional CPU: the PowerPC A2 chip has 16 in-order cores, 
each supporting 4 relatively light weight threads
and a crossbar on-chip network. A 17th core runs the OS functions and an 18th core is a spare to improve yields or take place of a damaged core. 
This design gives the PowerPC A2 chip similar performance to power ratio as an NVIDIA Tesla 2090 GPU, making Blue Gene/Q computers very efficient.
One of the main advantages of using GPUs (or GPU-like architectures) over traditional serial processors is the increased performance per watt
and the possibility to perform simulations on commodity hardware.

\section{Wave function methods}

The stochastic techniques based on Markov chains and the Metropolis algorithm showed great 
success in the field theory examples above. There are also many-body wave function 
methods that use the wave function as the central variable and use stochastic techniques 
for the actual numerical work. These quantum Monte Carlo (QMC) techniques have shown 
to be very powerful tools for studying electronic structures beyond the mean-field level of for example the density functional theory. A general overview of QMC can be found from \cite{qmc_rmp}. The simplest form of the QMC algorithms is the variational QMC, where a trial wave function with free 
parameters is constructed and the parameters are optimized, for example, to minimize the total energy \cite{sga}. This simple strategy works rather well for various different systems, even for
strongly interacting particles in an external magnetic field \cite{vmc_qd}.

There have been some works porting QMC methods to GPUs.
In the early work by Amos G. Anderson {\it et al.} \cite{qmc07}, the overall speedup compared to the CPU
was rather modest, from three to six, even if the individual kernels were up to 30 times faster.
More recently, Kenneth P. Esler {\it et al.} \cite{qmc12} have ported the QMCPack simulation code to the Nvidia CUDA GPU platform. Their full application speedups are typically around 10 to 15 compared to a quad-core Xeon CPU. This speedup is very promising and demonstrates 
the great potential GPU computing has for the QMC methods that are perhaps the computational technologies that are the mainstream in future electronic structure calculations.

There are also many-body wave function methods that are very close to the quantum chemical methods. One example of these is the full configuration interaction method in chemistry that is termed exact diagonalization (ED) in physics.
The activities in porting the quantum chemistry approaches to GPU are reviewed in \cite{Gotz}, and we try to remain on the physics side of this unclear borderline. We omit, for example, works on the coupled cluster method on the GPU \cite{coupled-cluster}. Furthermore, quantum mechanical transport problems are also not discussed here \cite{transport}.

Lattice models \cite{Hubbard,Gutzwiller} are important for providing a general understanding of many central physical concepts like magnetism. Furthermore, realistic materials can be cast to a lattice model \cite{qmc09}. Few-site models can be calculated exactly using the ED method. The ED method turns out to be very efficient on the GPU \cite{GPU_ED}.  
In the simplest form of ED, the first step is to construct the many-body basis 
and the Hamiltonian matrix in it. Then follows the most time-consuming part, namely the actual diagonalization
of the Hamiltonian matrix. In many cases, one is mainly interested in the lowest eigenstate and possibly a few of the lowest excited states. For these, the Lanczos algorithm turns out to be very suitable \cite{GPU_ED}. The basic idea of the Lanczos scheme is to map the huge but sparse Hamiltonian matrix to a smaller and tridiagonal form in the so-called Krylov space 
that is defined by the spanning vectors obtained from a starting vector $| x_0 \rangle$ by acting with the Hamiltonian as $H^m | x_0 \rangle$. Now, as the GPU is very powerful for the matrix-vector product, it is not surprising that high speedups compared to CPUs can be found\cite{GPU_ED}.

\section{Outlook}

The GPU has made a definite entry into the world of computational physics. Preliminary studies 
using emerging technologies will always be done, but the true litmus test of a new technology is whether
studies emerge where the new technology is actually used to advance science. The increasing frequency of 
studies that mention GPUs is a clear indicator of this.

From the point of view of high performance computing in computational physics, 
the biggest challenge facing GPUs at the moment is scaling: 
in the strong scaling case, as many levels of parallelism as possible inherent in the problem should be exploited
in order to reach the best performance with small local subsystems. 
The basic variables of the model are almost always
vectors of some sort, making them an ideal candidate for SIMD type parallelism. This is often achieved with CPUs
with a simple compiler flag, which instructs the compiler to look for opportunities to combine independent instructions
into vector operations. 

Furthermore, large and therefore interesting problems from a HPC point of view are typically 
composed of a large number of similar variables, be it particles, field values, cells or just entries in an array 
of numbers, which hints at another, higher level of parallelism of the problem that 
traditionally has been exploited using MPI, but is a prime candidate for a data parallel algorithm.
Also, algorithmic changes may be necessary to reach the best possible performance: 
it may very well be that the best algorithm for CPUs is no longer the best one for GPUs. A classic
example could be the question whether to use lookup tables of certain variables or
recompute them on-the-fly. Typically, on the GPU the flops are cheap making the recomputation an
attractive choice whereas the large caches of the CPU may make the lookup table a better option.

On the other hand, MPI communication latencies should be minimized and bandwidth
increased to accommodate the faster local solve to help with both weak and strong scaling. 
As far as we know, there are very few, if any, groups taking advantage of 
GPUDirect v.2 for NVIDIA GPUs \cite{gpudirect}, 
which allows direct GPU-to-GPU communications 
(the upcoming GPUDirect Support for RDMA will allow direct communications across network nodes)  
reducing overhead and CPU synchronization needs. 
Even GPUDirect v.1 helps,  as then one can share the \emph{pinned memory} buffers between 
Infiniband and GPU cards, removing the need to do extra local copies of data. 
The MPI implementations should also be scaled to fit the needs of the GPUs connected to the node:
currently the network bandwidth between nodes seems to be typically about two orders of magnitude lower
than the memory bandwidth from the GPU to the GPU memory, which poses a challenge to strong scaling,
limiting GPU applicability to situations with relatively large local problem sizes.

Another, perhaps an even greater challenge, facing GPUs and similar systems is the ecosystem: 
Currently a large portion of 
the developers and system administrators like to think of GPUs and similar solutions 
as \emph{accelerators} -- an accelerator is a component, which is attached to the main
processor and used to speed up certain portions of the code, but as these ``accelerators'' become
more and more agile with wider support for standard algorithms, the term becomes more and more
irrelevant as a major part of the entire computation can be done on the ``accelerator'' and the 
original ``brains'' of the machine, the CPU, is mainly left there to take care of administrative
functions, such as disk IO, common OS services and control flow of the program. 

As single threaded performance 
has reached a local limit, all types of processors are seeking more performance out of 
parallelism: more cores are added and vector units are broadened. This trend, fueled by the fact that 
transistor feature sizes keep on shrinking, hints at some type of convergence in the near future, 
but exactly what it will look like is anyone's best guess. 
At least in computational physics, it has been shown already that the scientists are willing 
to take extra effort in porting their code to take advantage of massively parallel architectures, 
which should allow them to do the same work with less energy and do more science
with the resources allocated to them.

The initial programming effort does raise a concern for productivity:
How much time and effort is one willing to spend to gain a certain amount of added performance? 
Obviously, the answer depends on the problem itself, but perhaps even more on the assumed direction 
of the industry -- a wrong choice may result in wasted effort if the chosen solution simply does not exist
in five years time. Fortunately, what seems to be clear at the moment, is the overall direction of the
industry towards higher parallelism, which means that a large portion of the work needed to parallelize
a code for a certain parallel architecture will most probably be applicable to another parallel architecture
as well, reducing the risk of parallelization beyond the typical MPI level.  

The answer to what kind of parallel architectures will prevail the current turmoil in the industry
may depend strongly on consumer behavior, since a large part of the development
costs of these machines are actually subsidized by the development of the consumer variants of the products.
Designing a processor only for the HPC market is too expensive and a successful product will need a sister
or at least a cousin in the consumer market. This brings us back to DOOM and other performance-hungry games: 
it may very well be that the technology developed for the gamers of today, will be the programming
platform for the scientists of tomorrow.

\subsection*{Acknowledgements}

We would like to thank Kari Rummukainen, Adam Foster, Risto Nieminen, Martti Puska, and
Ville Havu for all their support. Topi Siro acknowledges the financial support from the Finnish
Doctoral Programme in Computational Sciences FICS. 
This research has been partly supported by the Academy of
Finland through its Centres of Excellence Program (Project
No. 251748) and by the Academy of Finland Project No. 1134018.

\bibliographystyle{splncs}
%\bibliography{para12_physics}

\end{document}